\documentclass[12pt,a4paper]{article}
\usepackage[dvipdfm]{graphicx,color}
\usepackage{bm}
\usepackage{amsmath,amssymb}
\usepackage{booktabs}
\makeatletter
\def\Slash{\mathpalette\@Slash}
\def\@Slash#1#2{{\ooalign{\hfil$#1/$\hfil\crcr$#1{#2}$}}}
\makeatother
\begin{document}

\title{Nucleon decay via dimension 6 operators\\
in anomalous $U(1)_A$ SUSY GUT}
\author{
\centerline{
Nobuhiro~Maekawa$^{1,2}$\footnote{E-mail address: maekawa@eken.phys.nagoya-u.ac.jp}
~and 
Yu~Muramatsu$^{1}$\footnote{E-mail address: mura@eken.phys.nagoya-u.ac.jp}}
\\*[25pt]
\centerline{
\begin{minipage}{\linewidth}
\begin{center}
$^1${\it \normalsize Department of Physics, Nagoya University, Nagoya 464-8602, Japan }  \\*[10pt]
$^2${\it \normalsize Kobayashi Maskawa Institute, Nagoya University, Nagoya 464-8602, Japan }  \\*[10pt]\end{center}
\end{minipage}}
\\*[50pt]}
\date{}
\maketitle
\begin{abstract}
Nucleon lifetimes for various decay modes via dimension 6 operators 
are calculated in the anomalous $U(1)_A$ GUT scenario, in which the unification
scale $\Lambda_u$ becomes smaller than the usual supersymmetric (SUSY) unification scale
$\Lambda_G=2\times 10^{16}$ GeV in general. 
Since the predicted lifetime $\tau(p\rightarrow \pi^0+e^c)$ becomes
around the experimental lower bound though it is strongly dependent on the
explicit models, the discovery of the nucleon decay in near
future can be expected.
We explain why the two ratios 
$R_1\equiv\frac{\Gamma_{n \rightarrow \pi^0 + \nu^c}}{\Gamma_{p \rightarrow \pi^0 + e^c}}$ 
and $R_2\equiv\frac{\Gamma_{p \rightarrow K^0 + \mu^c}}{\Gamma_{p \rightarrow \pi^0 + e^c}}$ 
are important in identifying grand unification group and 
we show that three anomalous $U(1)_A$ SUSY GUT 
models with $SU(5)$, $SO(10)$ and $E_6$ 
grand unification group can be identified by measuring the two ratios.
If $R_1$ is larger than $0.4$, the grand unification group is not $SU(5)$, and moreover 
if $R_2$ is larger than $0.3$, the grand unification group is implied to be $E_6$.
\end{abstract}

\section{Introduction}
Grand unified theory (GUT)\cite{GUT} is one of the most promising 
possibilities among models beyond the standard model (SM). 
Theoretically, it can unify not only three gauge interactions into
a single gauge interaction but also quarks and leptons into
fewer multiplets. Moreover, experimentally, not only measured values of 
three gauge couplings in the SM can be explained quantitatively in
the supersymmetric (SUSY) GUTs, but also the various hierarchies
of masses and mixings of quarks and leptons can be understood qualitatively
by the unification of quarks and leptons in one generation into $\bf 10$ and
$\bf\bar 5$ of SU(5) if it is assumed that the $\bf 10$ matters induce
stronger hierarchies for Yukawa couplings than the $\bf\bar 5$ matters\cite{anarchy}.

One of the most important predictions in the GUTs is the nucleon decay
\cite{GUT,nucleon decay dim6,nucleondecayoperator,nucleon decay dim5}.
In general, GUTs require some new particles which are not included in the SM.
Some of these new particles induce the nucleon decay.
For example, the adjoint representation of SU(5) group has 24 dimensions, while
the sum of the dimensions for the adjoint representations of the SM gauge groups
is just 12. There are new gauge bosons in the SU(5) GUT, 
$X(\bf{\bar 3,2})_{\frac{5}{6}}$ and $\bar{X}(\bf{3,2})_{-\frac{5}{6}}$,
where $\bf \bar 3$ and $\bf 2$ mean anti-fundamental representation of 
$SU(3)_C$ and fundamental representation of $SU(2)_L$, respectively, 
 and $\frac{5}{6}$ is the hypercharge. 
These new gauge bosons induce dimension 6 effective operators which break both 
the baryon and lepton numbers and induce the nucleon decay. Usually, the main
decay mode of the proton via these dimension 6 operators is $p\rightarrow \pi^0 +e^c$. 
Since the mass of the superheavy gauge boson can be roughly estimated by the 
meeting scale of the three running gauge couplings, the lifetime of the nucleon
can be estimated in principle. Unfortunately, in the SM, three gauge couplings
do not meet at a scale exactly and the lifetime is proportional to the unification
scale to the fourth, and therefore, the prediction becomes in quite wide range.
However, if supersymmetry is introduced, the unification scale $\Lambda_G$ becomes
$2\times 10^{16}$ GeV, and therefore, the lifetime can be estimated as 
roughly $10^{36}$ years,
which is much larger than the experimental lower bound, $10^{34}$ years\cite{SuperK}. 

The partner of the SM doublet Higgs, which is called the triplet (colored) Higgs,
also induces the nucleon decay through Yukawa interactions. 
Since the Yukawa couplings for 
the first and second generation matters are much smaller than the gauge couplings, 
the constraint for the colored Higgs mass from the experimental limits of the 
nucleon decay is not so severe without SUSY. However, once SUSY is introduced, 
dimension 5 effective operators can break both baryon and lepton numbers and
induce nucleon decay\cite{nucleon decay dim5}. This can compensate the smallness of the Yukawa couplings.
Actually, in the minimal SU(5) SUSY GUT, the lower bound for the colored Higgs mass
becomes larger than the unification scale $\Lambda_G$
\cite{dim5bound,dim5bound'}. The experimental bound 
from the nucleon decay via
dimension 5 operators gives severe constraints for SUSY GUTs.

These constraints for the colored Higgs mass lead to the most difficult
problem in SUSY GUTs, i.e., the doublet-triplet splitting problem.
As noted above, the colored Higgs mass must be larger than the unification
scale, while the SM Higgs must be around the weak scale. 
Of course, a finetuning can realize such a large mass splitting even
in the minimal SUSY SU(5) GUT, but it is unnatural. In the literature, a lot of attempts 
have been proposed for this problem\cite{DTsplitting review}. However, in the most of the solutions, some terms
which are allowed by the symmetry are just neglected or the coefficients are taken to be
very small. Such requirements are, in a sense, finetuning, and therefore, some mechanism
is required which can realize such a large mass splitting in a natural way. 

Another famous problem in the SUSY GUTs is on the unrealistic Yukawa relations. 
The unification of matters results in the unification of the Yukawa couplings,
which often leads to unrealistic mass relations. 
In the minimal SU(5) GUT, 
the Yukawa matrix of the down-type quarks becomes the same as that of the charged 
leptons, which gives unrealistic predictions between masses of these particles.
In the minimal SO(10) GUT, all the Yukawa matrices become equivalent due to
the unification of all quarks and leptons in one generation into a single multiplet, $\bf 16$.
This Yukawa unification leads to unrealistic relations between masses of quarks 
and leptons.

It has been pointed out that 
if the anomalous $U(1)_A$ gauge symmetry is introduced, the doublet-triplet
splitting problem can be solved under the natural assumption that
all the interactions which are allowed by the symmetry of the theory are
introduced with $O(1)$ coefficients\cite{DTsplitting,GCUA,GCUA',E6LLM,E6Higgs}.
Note that the introduced interactions include
higher dimensional interactions. In the scenario, the nucleon decay via dimension
5 operators can be  strongly suppressed\cite{DTsplitting,GCUA}.
Moreover, with this natural assumption,
realistic quark and lepton masses and mixings can be obtained\cite{DTsplitting,E6LLM}. 
In this paper we denote such SUSY GUTs as the natural GUTs. 
One of the most interesting predictions of the natural GUTs is that 
the nucleon decay via dimension 6 operators is enhanced, i.e., the unification
scale $\Lambda_u$ becomes lower than $\Lambda_G$.
In the natural GUT, the unification scale is given as
\begin{equation}
\Lambda_u\sim \lambda^{-a}\Lambda_G,
\end{equation}
where $\lambda<1$ is the ratio of Fayet-Iliopoulos parameter to the cutoff $\Lambda$, which
is taken to be the usual SUSY GUT scale $\Lambda_G$ in the natural GUT in order to explain the
success of the gauge coupling unification\cite{GCUA,GCUA'}.
Since $a$ is the anomalous $U(1)_A$ charge of the adjoint Higgs and negative, 
the unification scale becomes smaller than the usual SUSY GUT scale.

In this paper, we study the nucleon decay via dimension 6 operators in the natural
GUTs. The grand unification group is $SU(5)$
\footnote{Strictly, in the literature, $SU(5)$ natural GUT has not been proposed.
However, we think that $SU(5)$ natural GUT is possible if the missing partner 
mechanism\cite{missing partner} is adopted.}
, $SO(10)$ or $E_6$.
In $SO(10)$ and $E_6$ unification we have additional gauge
bosons which induce nucleon decay in addition to the $X$ gauge bosons in $SU(5)$ GUT.
We will include these new effects due to the extra gauge bosons.
Moreover, we will include also the effects of the matrices which make Yukawa 
matrices diagonal.
The diagonalizing matrices are roughly fixed in the natural GUT 
in order to obtain the realistic quark and lepton mass matrices.
In the estimation, we will use the hadron matrix elements calculated by the 
lattice\cite{formfactor}.

\section{Decay widths of the nucleon}
In this section, we show how to estimate the partial decay widths of nucleon 
from the effective Lagrangian which induces nucleon decays.
The description in this section is based on the paper\cite{formfactor}.

In the standard model (SM), the dimension 6 operators which induce nucleon decay 
are classified completely\cite{nucleondecayoperator} and are written 
by one lepton $l$ and three quarks $q$ as
$\epsilon^{\alpha \beta \gamma}(\overline{l^c_{\Gamma}}q_{1\, \Gamma \, \alpha})
(\overline{q^c_{2\, \Gamma'}}_{\beta}q_{3\, \Gamma' \, \gamma})$
where $\alpha,\beta$, and $\gamma$ are color indices.
Here, $l^c$ is a charge conjugated field of the lepton $l$, and in this paper 
we denote $l^c_{\Gamma}$ as $(l_{\Gamma})^c$, where
the chirality indices $\Gamma, \Gamma'=L, R$.
In the following, color indices $\alpha,\beta$, and $\gamma$ are omitted in the operator 
$\epsilon^{\alpha \beta \gamma}(\overline{l^c_{\Gamma}}q_{1\, \Gamma \, \alpha})
(\overline{q^c_{2\, \Gamma'}}_{\beta}q_{3\, \Gamma' \, \gamma})$, i.e., we write 
it as 
$(\overline{l^c_{\Gamma}}q_{1\, \Gamma})(\overline{q^c_{2\, \Gamma'}}q_{3\, \Gamma'})$ 
for simplicity.
Once we calculate the effective Lagrangian which induces nucleon decays as
\begin{equation}
\mathcal{L}_{eff}=\sum _{I}C^I[(\overline{l^c_{\Gamma}}q_{1\, \Gamma})
(\overline{q^c_{2\, \Gamma'}}q_{3\, \Gamma'})]^I,
\end{equation}
where $C^I$ is a coefficient of the operator 
$[(\overline{l^c_{\Gamma}}q_{1\, \Gamma})(\overline{q^c_{2\, \Gamma'}}q_{3\, \Gamma'})]^I$, 
we can estimate the partial decay widths of the nucleon as follows.

In order to calculate the decay widths, we must know the hadron matrix elements 
with the initial nucleon state
$|N(\bm{k},s)\rangle$ with the momentum $\bm{k}$ and 
the spin $s$ and the final meson state $\langle \textrm{meson}(\bm{p})|$ with the 
momentum $\bm{p}$. These can be written as
\begin{equation}
\langle \textrm{meson}(\bm{p})|q_{1\, \Gamma}(\overline{q^c_{2\, \Gamma'}}q_{3\, \Gamma'})|
N(\bm{k},s)\rangle=
P_{\Gamma}[W_0^{\Gamma\, \Gamma'}(q^2)-
i\Slash{q}W_q^{\Gamma\, \Gamma'}(q^2)]u_N(\bm{k},s),
\label{eq:formfactor def}
\end{equation}
where $W_0^{\Gamma \, \Gamma'}, W_q^{\Gamma \, \Gamma'}$ are form factors and
$\bm{q}\equiv \bm{p}-\bm{k}$ is a momentum of the anti-lepton.
Here, $P_{\Gamma}$ is a chiral projection operator and $u_N(\bm{k},s)$ is a wave function
of the nucleon.
Usually, the first term in eq. (\ref{eq:formfactor def}) dominates over 
the second term because the anti-lepton is lighter than the nucleon.
Therefore, the hadron matrix elements
can be estimated as
\begin{equation}
\langle \textrm{meson}(\bm{p})|q_{1\, \Gamma}(\overline{q^c_{2\, \Gamma'}}q_{3\, \Gamma'})|
N(\bm{k},s)\rangle \simeq
P_{\Gamma}W_0^{\Gamma\, \Gamma'}(q^2)u_N(\bm{k},s).
\end{equation}
In our calculation, we use the form factor $W_0^{\Gamma\, \Gamma'}$ which has been calculated
 by Lattice \cite{formfactor} as in Table I. 
\begin{table}[tbp]
\begin{center}
\begin{tabular}{|c|c|c|} \hline
matrix element & $W_{0}^{RL},W_{0}^{LR}$ \\ \hline
$\langle \pi^0|(ud)u|p \rangle,\langle \pi^0|(du)d|n \rangle$ & -0.103(23)(34)  \\ \hline
$\langle \pi^+|(ud)d|p \rangle,-\langle \pi^-|(du)u|n \rangle$ & -0.146(33)(48)  \\ \hline
$\langle K^0|(us)u|p \rangle,-\langle K^-|(ds)d|n \rangle$   &  0.098(15)(12)  \\ \hline
$\langle K^+|(us)d|p \rangle,-\langle K^0|(ds)u|n \rangle$   & -0.054(11)(9)  \\ \hline
$\langle K^+|(ud)s|p \rangle,-\langle K^0|(du)s|n \rangle$   & -0.093(24)(18)  \\ \hline
$\langle K^+|(ds)u|p \rangle,-\langle K^0|(us)d|n \rangle$   & -0.044(12)(5)  \\ \hline
$\langle \eta|(ud)u|p \rangle,-\langle \eta|(du)d|n \rangle$   & 0.015(14)(17)  \\ \hline
\end{tabular}
\caption{Form factors for nucleon decays, which have been calculated by Lattice\cite{formfactor}.
First and second errors in $W_{0}^{RL},W_{0}^{LR}$ represent statistical and 
systematic ones, respectively.}
\end{center}
\end{table}

Then, we can estimate the partial decay widths for the process
$N\rightarrow \textrm{meson} +l^c$ as
\begin{align}
\Gamma(N\, \rightarrow \textrm{meson}\,+ l^{c})&=
\frac{1}{2m_N}\int 
\frac{d^3 p}{(2\pi)^3}\frac{1}{2E_p}
\frac{d^3 q}{(2\pi)^3}\frac{1}{2E_q}
\left| \mathcal{M}(m_N\rightarrow p+q) \right| ^2 \times \nonumber \\
&\qquad\qquad\qquad\qquad\qquad\qquad (2\pi)^4\delta ^{(4)}(k-p-q)  \nonumber \\
&\simeq  \frac{m_N}{32 \pi}
\left\{1- \left(\frac{m_{\textrm{meson}}}{m_N}\right)^2\right\}^2 \times \nonumber \\
&\qquad\qquad\left| \sum_{I} C^I W^I_0(N\, \rightarrow \textrm{meson})
\right| ^2,
\end{align}
where $m_N$ and $m_{\textrm{meson}}$ correspond to the masses of nucleons and mesons, respectively.
Partial lifetimes of the nucleon are defined as the inverse of the partial decay widths.

Therefore, once the coefficient $C^I$ is known, which is dependent on the concrete
models, the partial decay widths can be calculated.

\section{Calculation of the coefficient $C^I$}
In this section, we explain how to obtain the coefficients of the dim. 6 operators $C^I$
at the scale $\mu=m_N$. Firstly, we discuss the effective interactions which are induced
via superheavy gauge boson exchange. Secondly, we consider the effect of the unitary 
matrices which transform the flavor eigenstates to the mass eigenstates of quarks and
leptons. Finally, we calculate the renormalization factors by using the renormalization
group. 

The coefficients are strongly dependent on the explicit GUT models. 
Therefore, we have to fix GUT models which we consider in this paper.
First of all, we fix the grand unification group as $SU(5)$, $SO(10)$, or $E_6$, 
since the superheavy gauge bosons which induce the nucleon decay are dependent on 
the grand unification group.
We introduce $\bf 10$ of $SO(10)$ in addition to $\bf 16$ in $SO(10)$
GUT as matter fields. This is important in obtaining realistic quark and lepton masses and 
mixings in a natural way\cite{DTsplitting,GCUA}. 
(In $E_6$ GUT\cite{E6,SU(3)*3'}, the fundamental representation $\bf 27$ includes $\bf 10$ of 
$SO(10)$ as well as $\bf 16$.)
Moreover, we adopt the Cabibbo-Kobayashi-Maskawa (CKM) -like matrices and 
the Maki-Nakagawa-Sakata (MNS)-like matrices as the unitary matrices which transform 
flavor eigenstates to mass eigenstates of $\bf 10$ matters of $SU(5)$ and 
$\bf \bar{5}$ matters, respectively.

\subsection{Dim. 6 effective interactions via superheavy gauge boson exchange}
Before discussing the dim. 6 interactions which induce the nucleon decay, let us
recall how to unify the quarks and leptons in the SM into $E_6$ GUT multiplets,
because it is important to understand the embedding in $E_6$ GUTs in calculating
the nucleon decay and in grasping the 
meaning of the GUT models discussed in this paper.
All quarks and leptons are embedded into three $\bf 27$ multiplets of 
$E_6$. The fundamental representation $\bf 27$ is divided into several multiplets of
$SO(10)$ as
\begin{equation}
{\bf 27}\rightarrow {\bf 16} + {\bf 10} + {\bf 1}.
\end{equation}
The spinor $\bm{16}$ and the vector $\bm{10}$ of $SO(10)$ contain the SM multiplets as
\begin{equation}
{\bf 16}\rightarrow \underbrace{q_{L}({\bf 3,2})_{\frac{1}{6}}
+u^c_R({\bf \bar{3},1})_{-\frac{2}{3}}
+e^c_R({\bf 1,1})_1}_{{\bf 10}}+
\underbrace{d^c_R({\bf \bar{3},1})_{\frac{1}{3}}
+l_L({\bf 1,2})_{-\frac{1}{2}}}_{{\bf \bar{5}}}+
\underbrace{\nu^c_R({\bf 1,1})_0}_{{\bf 1}},
\end{equation}
\begin{equation}
{\bf 10}\rightarrow \underbrace{D^c_R({\bf \bar{3},1})_{\frac{1}{3}}
+L_L({\bf 1,2})_{-\frac{1}{2}}}_{{\bf \bar{5'}}}+
\underbrace{\overline{D^c_R}({\bf 3,1})_{-\frac{1}{3}}
+\overline{L_L}({\bf 1,2})_{\frac{1}{2}}}_{{\bf 5}},
\end{equation}
where the numbers denote the representations under the SM gauge group 
$SU(3)_C\times SU(2)_L\times U(1)_Y$.
Note that we have two $\bf \bar 5$ fields in one $\bf 27$. Therefore, if we introduce
three $\bf 27$s for three generations of quarks and leptons, we have six $\bf\bar 5$ fields
of $SU(5)$.
Three of six $\bf\bar 5$s become superheavy with three $\bf 5$ fields after breaking $E_6$ 
into the SM gauge group.
The other three
$\bf\bar 5$ fields and three $\bf 10$ of $SU(5)$ become quarks and leptons in three generations
in the SM. In this paper, $\bf\bar 5'$ denotes $\bf\bar 5$ fields from $\bf 10$ of $SO(10)$
to distinguish from $\bf\bar 5$ fields from $\bf 16$.
In the literature, it has been argued that the main components of matters 
in the SM come from the first and second generation ${\bf 27}_1$ and ${\bf 27}_2$ 
as $({\bf\bar 5}_1, {\bf\bar 5'}_1, {\bf\bar 5}_2)$\cite{E6LLM}
, which plays an important role in obtaining realistic 
quark and lepton masses and mixings. Here the index denotes the original flavor index 
for the $\bf 27$ of $E_6$. More details will be discussed in the next subsection.
Note that it is required to calculate the dim. 6 interactions for not only
usual unified fields $\bf 10$ and $\bf\bar 5$ fields but also $\bf\bar 5'$ fields
for $E_6$ GUT models.(Also in $SO(10)$ GUT models, the interactions which induce 
$\bf\bar 5'$ field must be calculated because we introduce $\bf{10}$ of $SO(10)$ 
as matter field.)

In $SU(5)$ GUTs, the superheavy gauge bosons for the nucleon decay are
$X$ and $\bar{X}$ which are included in the adjoint
gauge multiplet $\bf 24$ of $SU(5)$. 
Since $SU(5)$ is a subgroup of $SO(10)$ and $E_6$, the $X$ field contributes to nucleon decay 
even in $SO(10)$ and $E_6$ GUTs.
The $X$ field induces the effective dim. 6 interactions
which can be written in $SU(5)$ notation as 
$\bf(10_i^\dagger 10_i+\bar 5^\dagger_i\bar 5_i+\bar 5'^\dagger_i\bar 5'_i)
(10_j^\dagger 10_j+\bar 5^\dagger_j\bar 5_j+\bar 5'^\dagger_j\bar 5'_j)$
 where $\bf 10_i$ and $\bf\bar 5_i$ of $SU(5)$ are matter fields with flavor index $i,j$. Here, the terms including 
 $\bar 5'$ must be taken into account in $SO(10)$ or $E_6$
 GUT. 
In $SO(10)$ GUT models, additional fields
$X'$ and $\overline{X'}$ also induce nucleon decay. They are included in the adjoint gauge
field $\bf 45$ divided as
\begin{eqnarray}
{\bf 45}&\rightarrow& \underbrace{
G({\bf 8,1})_0+W({\bf 1,3})_0
+\overline{X}({\bf 3,2})_{-\frac{5}{6}}+X({\bf \bar{3},2})_{\frac{5}{6}}
+N^c({\bf 1,1})_0}_{{\bf24}} \nonumber \\
&+&\underbrace{X'({\bf 3,2})_{\frac{1}{6}}
+U'^c_R({\bf \bar{3},1})_{-\frac{2}{3}}
+E'^c_R({\bf 1,1})_1}_{{\bf 10}} \\
&+&\underbrace{\overline{X'}({\bf \bar{3},2})_{-\frac{1}{6}}
+\overline{U'^c_R}({\bf 3,1})_{\frac{2}{3}}
+\overline{E'^c_R}({\bf 1,1})_{-1}}_{{\bf \overline{10}}}+
\underbrace{N'^c({\bf 1,1})_0}_{{\bf 1}}. \nonumber
\end{eqnarray}
The effective interactions induced by $X'$ field are included 
in the effective interaction 
$(\bf 10^\dagger_i \bar 5_i)\cdot (\bf\bar 5^\dagger_j 10_j)$.  
Note that it does not include $\bf{\bar 5'}$ fields because the superfield $\bf 5$s 
are inevitable to appear in the effective interactions with $\bf{\bar 5'}$ fields.
In $E_6$ GUTs, the additional
superheavy gauge bosons $X''$ and $\overline{X''}$ can produce the nucleon decay.  The new
superheavy gauge bosons are included in $\bf 16$ and $\bf \overline{16}$ 
of $SO(10)$ in the adjoint $\bf 78$ of
$E_6$, which is divided as
\begin{equation}
{\bf 78\rightarrow 45 + 16 + \overline{16} + 1}.
\end{equation}
The $X''$ is included in $\bf 10$ of $SU(5)$ in $\bf 16$ and has the same quantum 
numbers as $X'$ under the SM gauge group.
This $X''$ field induces the effective interactions included in
$(\bf 10^\dagger_i \bar 5'_i)\cdot (\bf\bar 5'^\dagger_j 10_j)$.

By using the technique of decomposition of $E_6$ into the subgroup 
$SU(3)_C\times SU(3)_L\times SU(3)_R$\cite{SU(3)*3',SU(3)*3},
the dim. 6 effective interactions for quark and lepton flavor eigenstates can be 
calculated as
\begin{eqnarray}
\mathcal{L}_{eff}&=& \frac{g_{GUT}^2}{M_X^2}\left\{ \right.
(\overline{e^c_R}_i u_{Rj})(\overline{u_L^c}_j d_{Li})+
(\overline{e^c_R}_i u_{Rj})(\overline{u_L^c}_i d_{Lj}) \nonumber \\ &+&
(\overline{e^c_L}_i u_{Lj})(\overline{u_R^c}_j d_{Ri})+
(\overline{E^c_L}_i u_{Lj})(\overline{u_R^c}_j D_{Ri}) \nonumber \\ &-&
(\overline{\nu^c_L}_i d_{Lj})(\overline{u_R^c}_j d_{Ri})-
(\overline{N^c_L}_i d_{Lj})(\overline{u_R^c}_j D_{Ri}) \left. \right\}  \label{dim6int} \\ &+& 
\frac{g_{GUT}^2}{M_{X'}^2}\left\{
(\overline{e^c_L}_i u_{Lj})(\overline{u_R^c}_i d_{Rj})-
(\overline{\nu^c_L}_i d_{Lj})(\overline{u_R^c}_i d_{Rj})\right\} \nonumber \\ &+& 
\frac{g_{GUT}^2}{M_{X''}^2}\left\{
(\overline{E^c_L}_i u_{Lj})(\overline{u_R^c}_i D_{Rj})-
(\overline{N^c_L}_i d_{Lj})(\overline{u_R^c}_i D_{Rj})\right\}, \nonumber
\end{eqnarray}
where $g_{GUT}$ is the unified gauge coupling and 
the superheavy gauge boson masses $M_X$, $M_{X'}$, and $M_{X''}$ are dependent on the 
vacuum expectation values (VEVs) of the GUT Higgs which break $E_6$ into the SM gauge group. 
In this paper, we assume that the adjoint Higgs has the Dimopoulos-Wilczek (DW) 
type VEV\cite{DW form} as
\begin{equation}
\left< {\bf 45}_A \right>=i \sigma_2 \times \left(
\begin{array}{ccccc}
x & & & &  \\
 & x & & &  \\
 & & x & &  \\
 & & & 0 &  \\
 & & & & 0 
\end{array}
\right), \label{vev_form1}
\end{equation}
to solve the doublet-triplet splitting problem.
Here ${\bf 45}_A$ is the $\bf 45$ component field of the $E_6$ adjoint Higgs $A$
in $SO(10)$ decomposition
and $\sigma_i (i=1,2,3)$ is the Pauli matrix.
This is because in the anomalous $U(1)_A$ GUTs, the DW type VEV can be obtained in a natural
way and it is easier to obtain the realistic quark and lepton masses
and mixings than the other mechanism for solving the doublet-triplet splitting problem. This DW type VEV breaks $SO(10)$ into 
$SU(3)_C\times SU(2)_L\times SU(2)_R\times U(1)_{B-L}$.
The superheavy gauge boson masses are given by
\footnote{
Under $SU(3)_C\times SU(3)_L\times SU(3)_R$ decomposition of $E_6$, 
the gauge fields $\bar X$, $X'$, and $X''$ are included in $\bf(3,3,3)$ 
representation. Since $\bar X$ and $X'$ are $SU(2)_R$ doublet, 
the same contribution to the masses comes from the adjoint Higgs VEV $x$ which
 breaks $SO(10)$ into $SU(3)_C\times SU(2)_L\times SU(2)_R\times U(1)_{B-L}$.
From the fact that the $B-L$ charge is proportional to $\lambda_8^L+\lambda_8^R$,
 which are one of the generators of $SU(3)_L$ and $SU(3)_R$, we can calculate
 the contributions to the masses of $X$, $X'$, and $X''$ as in Eq. (\ref{Mass}).
Here $\lambda_A$ $(A=1,2,\cdots ,8)$ denotes the Gell-Mann matrices.
}
\begin{equation}
M_X^2=g_{GUT}^2 x^2,\quad
M_{X'}^2=g_{GUT}^2( x^2 + v_{c}^2), \quad
M_{X''}^2=g_{GUT}^2(\frac{1}{4} x^2 + v_{\phi} ^2).
\label{Mass}
\end{equation}
Here $v_{\phi}$ and $v_{c}$ are the VEV of the $E_6$ Higgs $\Phi({\bf 27})$ 
which breaks
$E_6$ into $SO(10)$ and the VEV of the $SO(10)$ Higgs $C({\bf 27})$ which breaks
$SO(10)$ into $SU(5)$, respectively. 
(And $\bar\Phi(\bf\overline{27})$ and $\bar C(\bf\overline{27})$ are also needed to
satisfy the $D$-flatness conditions of $E_6$.)
Note that the mass of the $X'$ gauge boson is almost the same as that of $X$ in 
anomalous $U(1)_A$ GUT because $v_c<<x$ in order to obtain the DW type VEV in a 
natural way\cite{DTsplitting,GCUA}.
In some of the typical $E_6$ GUTs with anomalous $U(1)_A$\cite{E6Higgs}, 
$v_{\phi}$ is smaller than $x$.
And therefore the $X''$ as well as the $X'$ can play an important role in nucleon decay.

Note that the interactions induced by $X''$ gauge boson are only between $\bf 10$ and
$\bf\bar 5'$ fields, 
while the interactions by $X'$ are only between $\bf 10$ and $\bf\bar 5$ fields and 
those by $X$ include various interactions among $\bf 10$, $\bf\bar 5$, and $\bf\bar 5'$.
Therefore, the $X''$ gauge boson contributes to the nucleon decay only for the restricted
models in which some of the first and second generation of quarks and leptons include the $\bf\bar 5'$ 
fields as the components. 

These VEVs can be
fixed by their anomalous $U(1)_A$ charges  as
\begin{equation}
x\sim \lambda^{-a}\Lambda ,v_c\sim \lambda^{-\frac{1}{2}(c+\bar{c})}\Lambda ,
v_{\phi}\sim \lambda^{-\frac{1}{2}(\phi+\bar{\phi})}\Lambda,
\end{equation}
where $a$, $\phi$, $\bar\phi$, $c$, and $\bar c$ are the anomalous $U(1)_A$ charges for
$A$, $\Phi$, $\bar\Phi$, $C$, and $\bar C$, respectively\cite{E6LLM}.
Each VEV has an $O(1)$ uncertainty that comes from $O(1)$ ambiguities in each term 
in the Lagrangian.
As mentioned in the introduction, the unification scale 
$\Lambda_u\equiv \langle A\rangle\sim x$ becomes lower than the usual GUT scale
$\Lambda_G\sim 2\times 10^{16}$ GeV because the cutoff $\Lambda=\Lambda_G$, 
$\lambda<1$, and the $U(1)_A$ charges for 
the Higgs fields like $A$ are negative in general. 
Therefore, the nucleon decay via dimension
6 operators is enhanced in the anomalous $U(1)_A$ GUT scenario\cite{GCUA}.
Here we consider two typical $U(1)_A$ charge assignments as
$(a=-1, \phi+\bar\phi=-1, c+\bar c=-4)$ and 
$(a=-1/2,\phi+\bar\phi=-2, c+\bar c=-5)$\cite{E6Higgs}.
In this paper we take $\lambda\sim 0.22$.
Note that relation $x>>v_c$ is always satisfied in the anomalous $U(1)_A$ GUT
with the DW type VEV, because the term which destabilizes the DW type VEV is 
allowed if $c+\bar c$ becomes larger.
It means that $X'$ gauge boson has sizable contribution to the nucleon decay
 in the anomalous $U(1)_A$ GUTs.
On the other hand, the relation $v_\phi>x$ is obtained in the former model, but
$v_\phi<x$ in the latter model. In this paper, we study the latter model because
$X''$ gauge boson has larger contribution to the nucleon decay. 
The prediction of the former model is similar to the $SO(10)$ model, because
the contribution of the $E_6$ gauge boson $X''$ becomes smaller.

The results in eq. (\ref{dim6int}) in $E_6$ GUT models can be applied to the
$SO(10)$ GUTs in the limit $M_{X''}\rightarrow \infty$ and to the $SU(5)$ GUTs in the limit
$M_{X''}, M_{X'}\rightarrow \infty$. If $\bf 10$ of $SO(10)$ is not introduced in $SO(10)$ 
models, just neglect the terms which include the $\bf\bar 5'$ fields in eq. (\ref{dim6int}).

\subsection{Realistic flavor mixings in anomalous $U(1)_A$ GUT models}
One of the most important features in the anomalous $U(1)_A$ models is that
the interactions can be determined by the anomalous $U(1)_A$ charges of the fields
except the $O(1)$ coefficients.
For example, 
the Yukawa interactions and the right-handed neutrino masses are
\begin{equation} 
Y_u^{ij}q_{Li}u^c_{Rj}h_u+Y_d^{ij}q_{Li}d^c_{Rj}h_d+Y_e^{ij}l_{Li}e^c_{Rj}h_d
+Y_{\nu_D}^{ij}l_{Li}\nu^c_{Rj}h_u+M_{\nu_R}^{ij}\nu^c_{Ri}\nu^c_{Rj},
\end{equation}
where the Yukawa matrices and the right-handed neutrino masses can be written by
\begin{eqnarray}
Y_u^{ij}&=&\lambda^{q_{Li}+u^c_{Rj}+h_u},\quad 
Y_d^{ij}=\lambda^{q_{Li}+d^c_{Rj}+h_d},\quad
Y_e^{ij}=\lambda^{l_{Li}+e^c_{Rj}+h_d},\nonumber \\
Y_{\nu_D}^{ij}&=&\lambda^{l_{Li}+\nu^c_{Rj}+h_u},\quad
M_{\nu_R}^{ij}=\lambda^{\nu^c_{Ri}+\nu^c_{Rj}}\Lambda
\end{eqnarray}
\cite{FNLLM}.
Here, $h_u$ and $h_d$ are the Higgs doublets for up quarks and for down quarks,
respectively.
We have used the notation in which the matter and Higgs fields and the minimal SUSY SM 
Higgs $h_u$ and $h_d$ and the 
$U(1)_A$ charges are written by the same characters as the corresponding fields.
By unitary transformation,
\begin{equation}
\psi'_{Li}=(L_{\psi}^{\dagger})_{ij}\psi_{Lj},\quad
\psi'^c_{Ri}=(R_{\psi}^{\dagger})_{ij}\psi^c_{Rj},
\end{equation}
where $\psi=u,d,e,\nu$,
these Yukawa matrices can be diagonalized. Since $u_L(\nu_L)$ and $d_L(e_L)$ are included in $q_L(l_L)$,
we use $q_L=u_L=d_L(l_L=\nu_L=e_L)$ as their $U(1)_A$ charges.
Here, $\psi'$ is a mass eigenstate and $\psi$ is a flavor eigenstate.
What is important in the anomalous $U(1)_A$ theory is that not only quark and lepton masses
but also the CKM matrix\cite{CKM} and the MNS matrix\cite{MNS}, which are defined as
\begin{equation}
U_{CKM}=L_u^{\dagger}L_d,\quad U_{MNS}=L_{\nu}^{\dagger}L_e,
\end{equation}
can be determined by their anomalous $U(1)_A$ charges as
\begin{eqnarray}
m_{u i}&=&\lambda^{q_{Li}+u^c_{Ri}+h_u}\langle h_u\rangle, \quad 
m_{d i}=\lambda^{q_{Li}+d^c_{Ri}+h_d}\langle h_d\rangle, \quad 
m_{e i}=\lambda^{l_{Li}+e^c_{Ri}+h_d}\langle h_d\rangle, \nonumber \\
m_{\nu_i}&=&\lambda^{2l_{Li}+2h_u}\frac{\langle h_u\rangle^2}{\Lambda},\quad
(U_{CKM})_{ij}=\lambda^{|q_{Li}-q_{Lj}|}, \quad 
(U_{MNS})_{ij}=\lambda^{|l_{Li}-l_{Lj}|}, 
\end{eqnarray}
except $O(1)$ coefficients. 
Any mass hierarchies can be obtained by choosing the appropriate $U(1)_A$ charges,
but we have several simple predictions for mixings as
$(U_{CKM})_{13}\sim (U_{CKM})_{12}(U_{CKM})_{23}$,
$(U_{MNS})_{13}\sim (U_{MNS})_{12}(U_{MNS})_{23}$,
$(U_{MNS})_{23}^4\sim (m_{\nu 3}^2-m_{\nu 2}^2)/(m_{\nu 2}^2-m_{\nu 1}^2)$.
Note that normal hierarchy for neutrino masses is also predicted.
Not only these predictions are consistent with the observations but also
realistic quark and lepton masses and mixings can be obtained
by choosing the $U(1)_A$ charges. For example, if we take $q_{L1}-q_{L2}=1$ and
$q_{L2}-q_{L3}=2$, we can obtain the realistic CKM matrix when $\lambda\sim 0.22$.
Taking $l_{L1}\sim l_{L2}\sim l_{L3}$, the neutrino mixings become large.

In the $SU(5)$ unification, because of the unification of matters, 
we have some constraints among their $U(1)_A$ charges as 
$a_i\equiv q_{Li}=u^c_{Ri}=e^c_{Ri}$ and $\bar f_i\equiv d^c_{Ri}=l_{Li}$.
Then basically these charges are fixed in order to obtain realistic quark and 
lepton mixings. 
It is quite impressive that even with these charges, realistic hierarchical structures of quark and lepton masses are also obtained.
Actually the requirement results in that up type quarks have the largest mass hierarchy, 
neutrinos have the weakest, and down-type quarks and the charged leptons have middle 
mass hierarchies. These are nothing but the observed mass hierarchies for quarks and 
leptons, though the first generation neutrino mass has not been observed yet. 
The unrealistic GUT relation $Y_d=Y_e^t$ can be easily avoided in anomalous $U(1)_A$
GUT because the higher dimensional interaction 
$\lambda^{a_i+\bar f_j+a+h_d}A_iA\bar F_jH_d$, which breaks the unrealistic GUT 
relation $Y_d=Y_e^t$ after developing the VEV of the adjoint Higgs $A$ as 
$\langle A \rangle\sim\lambda^{-a}$, gives the same order contribution to the Yukawa
couplings as the original Yukawa interactions 
$\lambda^{a_i+\bar f_j+h_d}A_i\bar F_jH_d$. Here $A_i$ is $\bf 10$ matter of $SU(5)$
and $\bar F_i$ is $\bf\bar 5$ matter.

However, in the minimal $SO(10)$ GUT, all quarks and leptons in one generation can be unified
into a single multiplet, and therefore, the $U(1)_A$ charges for $q_L$ become the same
as $l_L$. That leads to the same mixings of quarks and leptons.
This is unrealistic prediction in the minimal
$SO(10)$ GUT with the anomalous $U(1)_A$ gauge symmetry.

There are several solutions to realize realistic flavor mixings 
in $SO(10)$ GUT models. 
One of them is by introducing one or a few additional $\bm{10}$ of $SO(10)$ 
 fields as matter fields.
 When one of the $\bf\bar 5$ fields of $SU(5)$ from 
 $\bf 10$ of $SO(10)$ becomes quarks and leptons, 
 we have different $U(1)_A$ charge hierarchy for $\bf\bar 5$ fields from that for
 $\bf 10$ of $SU(5)$.
 As the result, the realistic quark and lepton masses and
 mixings can be obtained\cite{DTsplitting,GCUA}.
 
One of the most important features in $E_6$ unification is that the additional 
$\bf 10$ of $SO(10)$ fields in $SO(10)$ unification are automatically introduced,
 because the fundamental representation $\bf 27$ includes $\bf 10$ in addition to
  $\bf 16$ of $SO(10)$. Moreover, the assumption in $SU(5)$ unification that
the $\bf 10$ fields induce stronger hierarchical Yukawa couplings than
$\bf\bar 5$ fields can be derived in $E_6$ unification. 
Since three $\bf 27$ matters are introduced for the 
quarks and leptons, we have six $\bf\bar 5$ fields. Three of six 
$\bf\bar 5$ fields become superheavy with three $\bf 5$ fields as noted in the previous 
subsection. 
Since the third generation field $\bf 27_3$ has larger Yukawa couplings due to 
smaller $U(1)_A$ charge, it is natural that two $\bf\bar 5$ fields from $\bf 27_3$
become superheavy, and therefore, three massless $\bf\bar 5$ fields come from
the first and second generation fields $\bf 27_1$ and $\bf 27_2$\cite{E6LLM}.  
As the result, we can obtain milder hierarchy for $\bf\bar 5$ fields than the
original hierarchy for $\bf 10$ fields, which is nothing but what we would like
to explain. Main modes of typical three massless $\bf\bar 5$ fields are 
$\bf\bar 5_1$, $\bf\bar 5_1'$, and $\bf\bar 5_2$. 
It is important that the Yukawa couplings of $\bf\bar 5'$ can also be controlled by 
$U(1)_A$ charges of matters and Higgs.
Therefore we can choose which $\bf\bar 5$ field becomes $\bf\bar 5_1'$ 
by fixing the $U(1)_A$ charges.
In order to obtain the larger neutrino mixings, it is the best that the main 
component of the second generation $\bf\bar 5$ is $\bf\bar 5_1'$. 
Though we do not discuss here the details for the realistic models and explicit charge  assignments, we can obtain realistic mixing matrices as 

\begin{equation}
U_{CKM}=\begin{pmatrix}
1 & \lambda & \lambda^3 \\
\lambda & 1 & \lambda^2 \\
\lambda^3 & \lambda^2 & 1
\end{pmatrix},U_{MNS}=\begin{pmatrix}
1 & \lambda^{\frac{1}{2}} & \lambda \\
\lambda^{\frac{1}{2}} & 1 & \lambda^{\frac{1}{2}} \\
\lambda & \lambda^{\frac{1}{2}} & 1
\end{pmatrix}.
\end{equation}
These matrices have $O(1)$ uncertainties which come from $O(1)$ ambiguities 
of Yukawa interactions.

For the calculation of the nucleon decay widths, the explicit flavor structure
is quite important. Strictly speaking, these massless modes have mixings with the superheavy 
fields $\bf\bar 5_2'$, $\bf\bar 5_3'$, and $\bf\bar 5_3$, but in our calculation, 
we just neglect these mixings because their contribution is quite small.
We just consider the mixings between 
$\bf\bar 5_1$ and $\bf\bar 5_1'$, and $\bf\bar 5_2$ in $E_6$ unification.

\subsection{Renormalization factor}
To calculate coefficients for the dim. 6 effective interactions at the nucleon mass
scale, we have to consider the renormalization factors.
For the calculation, we have to divide the scale region into two parts.
The first region is from the GeV scale to the SUSY breaking scale. 
We call the effect from this region "long distance effect" and 
the renormalization group factor
is written as $A_{Rl}$\cite{RFL}.
The other region is from the SUSY scale to the GUT scale.
We call the effect from this region "short distance effect" and the renormalization
group factor is written as $A_{Rs}$\cite{RFS1,RFS2}.
The total renormalization factor $A_R$ is defined as followed:
\begin{equation}
A_R = A_{Rl} \times A_{Rs}.
\end{equation}
To calculate coefficients of dim. 6 effective interactions at the GeV scale, 
we multiply the renormalization factor by the dim. 6 effective interactions 
at the GUT scale.

One loop calculation gives the renormalization factor for each region and for 
each gauge interaction as
\begin{equation}
A_{Ri}=\left(\frac{\alpha_i (M_{end})}{\alpha_i (M_{start})} 
\right)^{\frac{A_i}{b_i}},
\end{equation}
\begin{equation}
\gamma_i = -2A_i\frac{g_i^2}{(4\pi)^2},\quad
\beta_i = b_i\frac{g_i^3}{(4\pi)^2},
\end{equation}
where $\gamma_i$ is the anomalous dimension for dim. 6 operators for each SM 
gauge interaction and 
$\beta_i$ is the $\beta$ function for each gauge coupling.
$M_{start}$ and $M_{end}$ are the energy scale of the boundary of each
region. ($M_{end}>M_{start}$.)

The value is dependent on the explicit GUT model. 
In this paper, we use the renormalization factor of the minimal SUSY $SU(5)$ GUT as
$A_R=3.6$,
for the dimension 6 operators which include a right-handed charged lepton $e_R^c$ 
and $A_R=3.4$ for the operators which include the doublet leptons $l$ 
as the reference values\cite{RFS2}.
In order to apply our results to an explicit GUT model,
the correction for the renormalization factor is needed.
For example, in an anomalous $U(1)_A$ SUSY $SO(10)$ GUT
(explicit $U(1)_A$ charges are given in Figure caption of Fig.1 in Ref\cite{GCUA}), 
the renormalization factor can be estimated as
$A_R=3.2$ $(A_{Rl}=1.5,A_{Rs}=2.1)$
for the operators which include the singlet charged lepton $e_R^c$.
In this model, the gauge couplings become larger because there are a lot of
 superheavy particles, which increases the renormalization factor. However, 
the unification scale is lower, which decreases the renormalization factor.
The latter effect is larger in this model.
Therefore, the nucleon lifetime in this anomalous $U(1)_A$ SUSY $SO(10)$ GUT model 
\footnote{
Strictly, the absolute value of the adjoint Higgs VEV $\langle A \rangle$ in 
this model is different from the VEVs adopted in this paper.
The correction about $(2)^{-4}$ is needed for the lifetime of nucleon.} is
$(3.2/3.6)^{-2}=1.3$ times longer than the calculated values by using the 
renormalization group factor in the minimal $SU(5)$ SUSY GUT model.

\section{GUT models}
In order to obtain realistic quark and lepton masses and mixings in anomalous 
$U(1)_A$ GUT scenario, the diagonalizing matrices for $\bf\bar 5$ fields have
large mixings as MNS matrix while those for $\bf 10$ fields have small mixings
as CKM matrix. Namely, 
\begin{eqnarray}
L_u&\sim & L_d\sim R_u\sim R_e\sim U_{CKM}, \\
R_d&\sim & L_e\sim L_\nu\sim U_{MNS}.
\label{diagonalizing}
\end{eqnarray}
Since the quark and lepton mixings are determined by the charges of left-handed quarks and 
leptons, respectively, 
the above result for diagonalizing matrices is inevitable in the anomalous $U(1)_A$
GUT.

We calculate various nucleon decay modes in the following anomalous $U(1)_A$ GUT models.
\begin{enumerate}
\item{\bf $SU(5)$ Model}

In $SU(5)$ unification, without loss of generality, we can take one of the
diagonalizing matrices for $\bf 10$ fields and one of the
diagonalizing matrices for $\bf\bar 5$ fields as unit matrices by
field redefinitions. In this paper, we take $R_u=1$ and $R_d=1$. 
Because of the relations $U_{CKM}=L_u^\dagger L_d$ and $U_{MNS}=L_\nu^\dagger L_e$,
we have three independent diagonalizing matrices in $SU(5)$ unification.
\item{\bf $SO(10)$ Model 1}

In $SO(10)$ unification, one  $\bf 10$ of $SO(10)$ is introduced as
 additional matter fields in order to obtain realistic quark and lepton masses 
 and mixings. It is essential that since $\bf\bar 5_3$ becomes superheavy with 
 $\bf 5$ and is replaced with the $\bf\bar 5'$ from the additional fields, the diagonalizing matrices for $\bf\bar 5$ fields can be much different from
 $\bf 10$ of $SU(5)$ fields. 
Note that the main modes of $\bf\bar 5$ fields become
$(\bf\bar 5_1, \bar 5', \bar 5_2)$. It is reasonable that the $\bf\bar 5'$  
becomes
the second generation $\bf\bar 5$ field to obtain the large neutrino
mixings. 
Without loss of generality, we can take one of the
 diagonalizing matrices as a unit matrix, and in this paper, we take 
 $R_u=1$. 
Because of the relations $U_{CKM}=L_u^\dagger L_d$ and $U_{MNS}=L_\nu^\dagger L_e$,
we have four independent diagonalizing matrices in $SO(10)$ unification.
\item{\bf $E_6$ Model 1}

In $E_6$ unification, the additional $\bf 10$ of $SO(10)$ matters are included in
the fundamental representation $\bf 27$ of $E_6$ in addition to $\bf 16$ .
It is reasonable that $\bf\bar 5$ fields from $\bf 27_3$ become 
superheavy because they have larger couplings than $\bf 27_1$ and $\bf 27_2$. Therefore, $\bf\bar 5$ fields 
in the standard model come from $\bf 27_1$ and 
$\bf 27_2$. The main modes become $(\bf\bar 5_1, \bar 5_1', \bar 5_2)$. 
If the $\bf\bar 5_1'$  becomes the second generation $\bf\bar 5$ field, 
the large neutrino mixings can be obtained as noted in the previous section.
We have four independent diagonalizing matrices as in $SO(10)$ unification.
\end{enumerate}

The values of the GUT Higgs VEVs are also important to calculate the partial
decay widths of nucleon. In the anomalous $U(1)_A$ GUT, these are fixed by
their $U(1)_A$ charges.
In these models, we take these VEVs as
\begin{equation}
x=1\times 10^{16} \text{GeV},\quad v_c=5\times 10^{14} \text{GeV},\quad 
v_\phi=5\times 10^{15} \text{GeV}.
\end{equation}
We have two typical $U(1)_A$ charge assignments in $E_6$ unification which give
$(x\sim\lambda^{0.5}\Lambda_G, v_c\sim\lambda^{2.5}\Lambda_G, v_\phi\sim\lambda\Lambda_G)$ and 
$(x\sim\lambda\Lambda_G, v_c\sim\lambda^{2}\Lambda_G, v_\phi\sim\lambda^{0.5}\Lambda_G)$.  
We adopted the former assignment in these models because the contribution from 
$E_6$ gauge boson $X''$ becomes larger. The latter assignment gives the similar
results as in $SO(10)$ model. 

\section{Numerical calculation}
In our calculation, the ambiguities in the diagonalizing matrices are considered
by randomly generating ten unitary matrices for each independent $L_f$ and 
$R_f$ ($f=u,d,e,\nu$). 
The unitary matrices must satisfy the following requirements:
\begin{enumerate}
\item We take real unitary matrices for simplicity.
\item $L_u=L_dU_{CKM}^{(exp)\dagger}$ and $L_\nu=L_eU_{MNS}^{(exp)\dagger}$ where
\begin{equation}
U_{CKM}^{(exp)}=\begin{pmatrix}
0.97 & 0.23 & 0.0035 \\
-0.23 & 0.97 & 0.041 \\
0.0086 & -0.040 & 1.0
\end{pmatrix},
U_{MNS}^{(exp)}=\begin{pmatrix}
0.83 & 0.54 & 0.15 \\
-0.48 & 0.53 & 0.70 \\
0.30 & -0.65 & 0.70
\end{pmatrix}
\end{equation}
\cite{PDG,PDGc,theta13}.
\item $L_u\sim L_d\sim R_e\sim U_{CKM}$ and $L_\nu\sim L_e(\sim R_d)\sim U_{MNS}$
where
\begin{equation}
U_{CKM}=\begin{pmatrix}
1 & \lambda & \lambda^3 \\
\lambda & 1 & \lambda^2 \\
\lambda^3 & \lambda^2 & 1
\end{pmatrix},\quad
U_{MNS}=\begin{pmatrix}
1 & \lambda^{0.5} & \lambda \\
\lambda^{0.5} & 1 & \lambda^{0.5} \\
\lambda & \lambda^{0.5} & 1
\end{pmatrix}.
\end{equation}
Each component has $O(1)$ coefficient $C_{ij}$, and we take
$0.5\leq C_{ij}\leq 2$.
\end{enumerate}
Since we have three independent diagonalizing matrices in $SU(5)$ unification, 
we examine $10^3$ model points. In $SO(10)$ and $E_6$ unification, four independent
diagonalizing matrices lead to $10^4$ model points.

\subsection{Various decay modes for proton}
\begin{figure}
 \centering
 \includegraphics[width=12cm,clip]{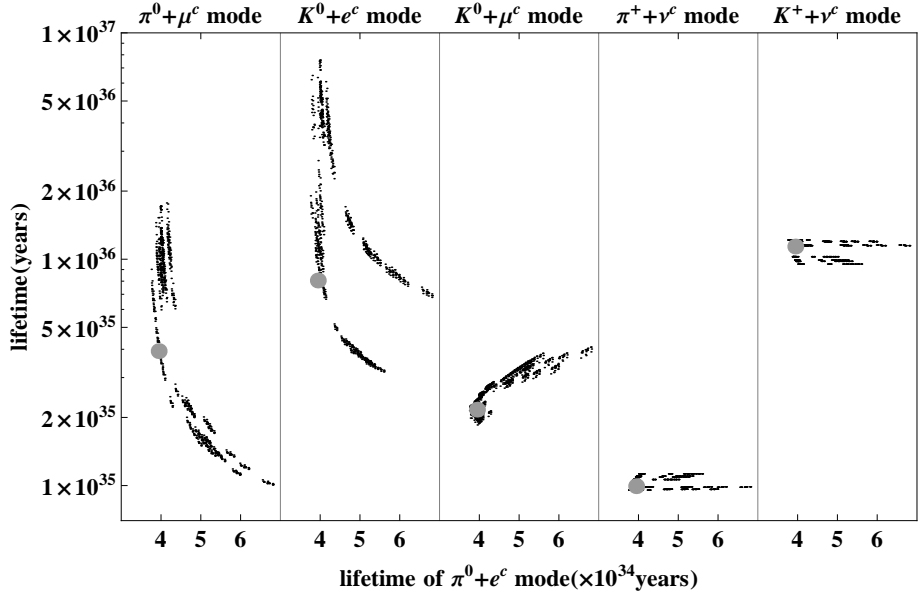}
 \caption{Various lifetimes of proton in $SU(5)$ Model with $M_X=g_{GUT}x$ and 
 $x=1\times 10^{16}\text{GeV}$.
 The gray large circles show the predictions of the minimal $SU(5)$ GUT model\cite{dim5bound'}.}
 \label{FigSU5}
\end{figure}

\begin{figure}
 \centering
 \includegraphics[width=12cm,clip]{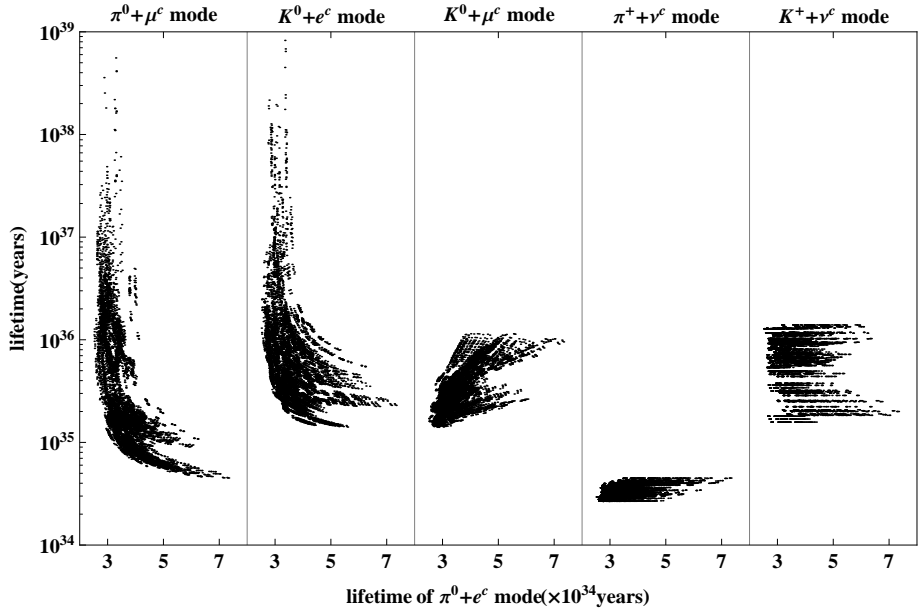}
 \caption{Various lifetimes of proton in $SO(10)$ Model 1 with $M_X=g_{GUT}x$, 
 $M_{X'}=g_{GUT}\sqrt{x^2+v_c^2}$, $x=1\times 10^{16}\text{GeV}$, and $v_c=5\times 10^{14}\text{GeV}$.}
 \label{FigSO10}
\end{figure}

\begin{figure}
 \centering
 \includegraphics[width=12cm,clip]{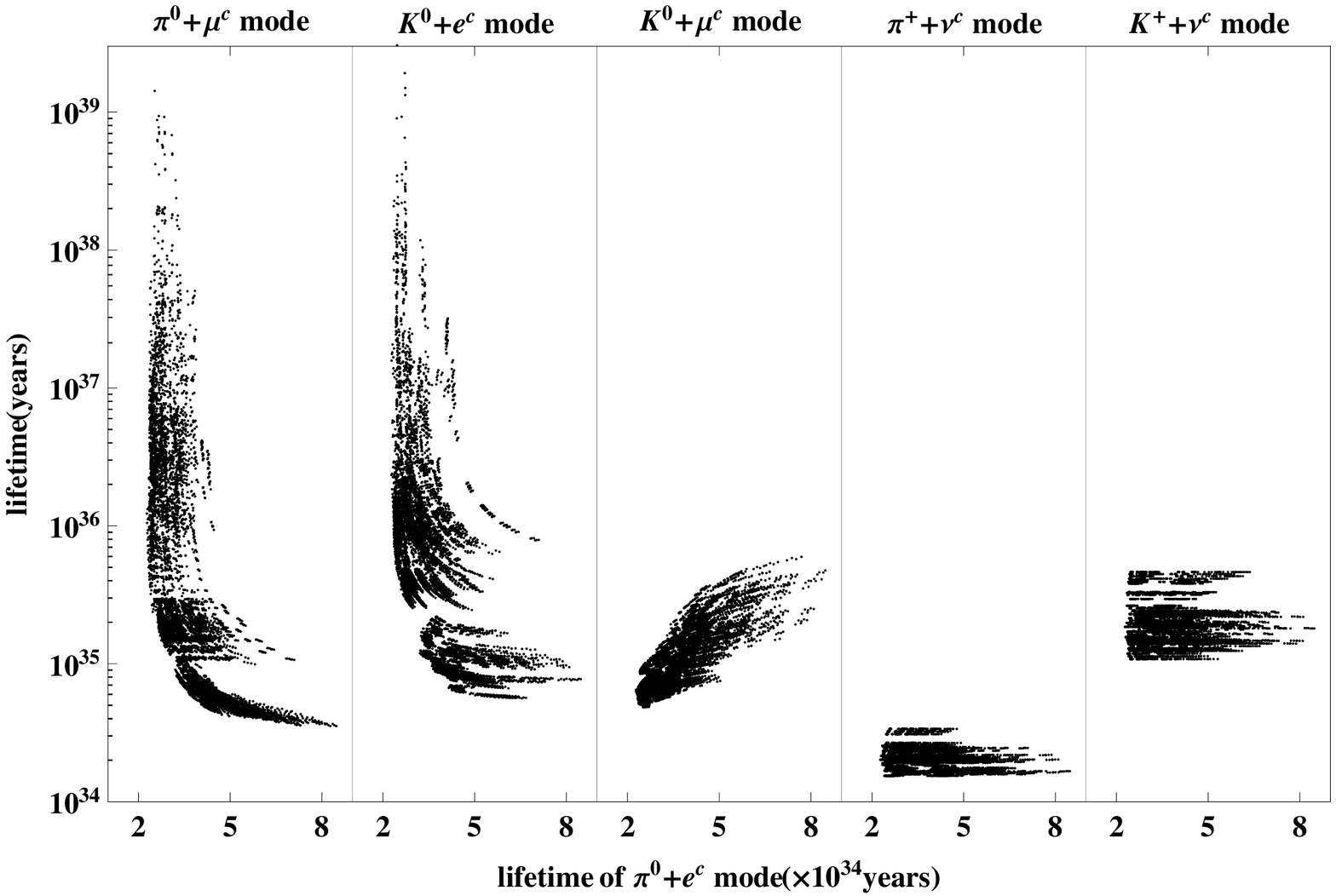}
 \caption{Various lifetimes of proton in $E_6$ Model 1 with $M_X=g_{GUT}x$, 
 $M_{X'}=g_{GUT}\sqrt{x^2+v_c^2}$, $M_{X''}=g_{GUT}\sqrt{\frac{x^2}{4}+v_{\phi}^2}$, 
 $x=1\times 10^{16}\text{GeV}$, $v_c=5\times 10^{14}\text{GeV}$, and $v_c=5\times 10^{15}\text{GeV}$.}
 \label{FigE6}
\end{figure}

We calculate the lifetime of the proton for various decay modes.
The results are shown in Figure  \ref{FigSU5}, \ref{FigSO10} and \ref{FigE6}.
We plot the lifetime of the most important decay mode, $p\rightarrow\pi^0+e^c$,
on the horizontal axis and the lifetime of the other decay modes on the
vertical axis. 
In Figure  \ref{FigSU5} the gray large circles show the predictions of the minimal 
$SU(5)$ GUT model
in which all the diagonalizing matrices can be fixed\cite{dim5bound'}, although it has unrealistic
 GUT relations for the Yukawa couplings between the charged leptons and the 
 down-type quarks.
Here, we used the same value for the VEV $x$ as the value we adopted in this paper.
 
We have several comments on these results.
First, the predicted lifetime of $p\rightarrow \pi^0 + e^c$ decay mode is not 
far from the experimental lower bound, $\tau(p\rightarrow \pi^0 + e^c)>1.29\times 10^{34}$
years\cite{SuperK}. Note that these results are obtained for
the models with the unification scale $\Lambda_u\sim 1\times 10^{16}$ GeV.
Therefore, for the models with $a=-1$ (typically $\Lambda_u\sim 5\times 10^{15}$ GeV),
 the predicted
value becomes more than one order shorter. Of course, since we have the $O(1)$ 
ambiguity for the unification scale, which easily leads to more than one order 
longer predicted lifetime, and the hadron matrix elements have still
large uncertainties, these models ($a=-1$) cannot be excluded by this observation.
What is important here is that we should not be surprised if the nucleon
 decay via dim. 6 operators will be observed in very near future.
Second, the lifetimes of the decay modes which include an anti-neutrino 
are calculated by summing up the partial decay widths for different anti-neutrino
 flavor because the flavor of the neutrino cannot be distinguished by the present
experiments for nucleon decay. As the result, the lifetime of the decay modes which
include an anti-neutrino have less dependence on the parameters because
the dependence can be cancelled due to unitarity of the diagonalizing matrix 
$L_\nu$\cite{neutrinoSUM}.
Third, the flavor changing decay modes, for example, $p\rightarrow \pi^0 + \mu^c$ 
and $p\rightarrow K^0 + e^c$ decay modes, have stronger dependence on the explicit
$O(1)$ parameters in the diagonalizing matrices than the flavor unchanging decay
modes, $p\rightarrow \pi^0 + e^c$ and $p\rightarrow K^0 + \mu^c$ decay modes.
This is mainly because off-diagonal elements have stronger ambiguities than the
diagonal elements in diagonalizing matrices. 
Forth, we comment on the shape for the $p\rightarrow \pi^0 + \mu^c$, 
$p\rightarrow K^0 + e^c$, and  $p\rightarrow K^0 + \mu^c$ modes. 
Because of the unitarity of $L_e$ and $R_e$, 
the longer lifetime of $p\rightarrow \pi^0 + e^c$ leads to the shorter lifetime
of $p\rightarrow \pi^0 + \mu^c$ and $p\rightarrow K^0 + e^c$ modes and the longer
lifetime of $p\rightarrow K^0 + \mu^c$ mode. These tendencies can be seen in 
the figures. 

Finally, we comment on the shape of the figure for the decay modes which include 
an anti-neutrino. 
In the figures, a lot of lines which parallel the horizontal axis can be seen.
This is because the $O(1)$ parameters in the diagonalizing matrices, 
$L_e$ and $R_e$, change the lifetime of $p\rightarrow \pi^0 + e^c$ decay mode, but do not 
change the lifetime of decay modes which have anti-neutrino in the final state.
$L_e$ would change the lifetime of decay modes with anti-neutrino through the relation 
$L_{\nu}=L_e U_{MNS}^{(exp)\dagger}$.
However, as noted above, the different $L_{\nu}$s have the same contribution to the 
decay modes with anti-neutrino in which all different flavors are summed up, 
because of the unitarity of $L_{\nu}$.

In the next subsection, we would like to discuss how to identify the GUT models
by the nucleon decay modes. For the identification, we use 
$p\rightarrow \pi^0 + e^c$, $n\rightarrow \pi^0 + \nu^c$, and 
$p\rightarrow K^0+\mu^c$ decay modes because these are less dependent on the 
$O(1)$ parameters, where $n\rightarrow \pi^0 + \nu^c$ mode has also only small 
dependence on the $O(1)$ parameters as the $p\rightarrow \pi^+ + \nu^c$ has.

\subsection{Identification of GUT models}
In this subsection, we discuss how to distinguish GUT models by the nucleon decay.
We emphasize that the ratios of the partial decay widths for 
$p\rightarrow \pi^0 + e^c$, $n\rightarrow \pi^0 + \nu^c$, and 
$p\rightarrow K^0+\mu^c$ are important for the identification of GUT models.
The partial decay width is strongly dependent on the explicit values of the VEVs.
However, by taking the ratio, part of the dependence can be cancelled.
The results become independent of the absolute magnitudes of these VEVs and 
are dependent only on the ratios of the VEVs.
Therefore, the results can be applied to other GUT models with different VEVs, 
but with the same ratios of VEVs.

First, we would like to explain that  
the ratio of decay width for $n\rightarrow \pi^0 + \nu^c$ mode to 
decay width for $p\rightarrow \pi^0 + e^c$ mode 
is useful to distinguish GUT models\cite{SU5 VS SO10}, especially the grand unification group.
In $SU(5)$ GUT models as in eq. (\ref{dim6int}) there are four effective interactions which are important
for the nucleon decay. Three of them
induce the decay modes which include $e^c$ in the final state, while just one of them
causes the decay modes which include $\nu^c$. Therefore, in $SU(5)$ unification,
the ratio  becomes quite smaller than 1.
In $SO(10)$ unification, two effective interactions are added, which contribute
to the decay modes with $e^c$ and to those with $\nu^c$ equivalently.
In $E_6$ unification, two effective interactions with $E^c$ and with $N^c$ are added, 
and the contribution to $n\rightarrow \pi^0 + \nu^c$ through the flavor mixings 
becomes larger than the contribution to $p\rightarrow \pi^0 + e^c$.
Here the essential point is that the $SO(10)$ superheavy gauge boson $X'$ and the $E_6$ superheavy
gauge boson $X''$ induce only the effective interactions which include $\bf\bar 5$
fields of $SU(5)$ while the $SU(5)$ superheavy gauge boson $X$ can induce also
the effective interactions which include only $\bf 10$ of $SU(5)$.
Therefore, basically, the models with the larger grand unification group lead to the larger
ratio if the contributions from $X'$ and $X''$ are not negligible. 
This feature is useful to identify the grand unification group, especially when
the $X'$ and $X''$ are as light as the X. In the anomalous $U(1)_A$ GUT models,
the masses of $X'$ and $X''$ can be comparable to the $X$ mass, or even smaller
than the mass of $X$. Therefore, this identification is quite useful.

We calculate the ratio of decay width for $p\rightarrow \pi^0 + e^c$ mode to 
decay width for $n\rightarrow \pi^0 + \nu^c$ mode for the anomalous $U(1)_A$ GUT
models as
\begin{eqnarray}
R_1&\equiv&\frac{\Gamma_{n \rightarrow \pi^0 + \nu^c}}{\Gamma_{p \rightarrow \pi^0 + e^c}} =
\begin{cases}
0.18 - 0.34 & \text{$SU(5)$ Model} \\ 
0.35 - 0.90 & \text{$SO(10)$ Model 1} \\
0.38 - 2.5 & \text{$E_6$ Model 1}
\end{cases}.
\end{eqnarray}
It is obvious that 
the ratio
$\frac{\Gamma_{n \rightarrow \pi^0 + \nu^c}}{\Gamma_{p \rightarrow \pi^0 + e^c}}$ 
becomes larger for the larger grand unification group.
However, we cannot distinguish these GUT models by this ratio perfectly because 
we have the $O(1)$ ambiguities in the diagonalizing matrices. 
There is a region in which both $SO(10)$ and $E_6$ GUTs are allowed.

\begin{figure}[htb]
 \centering
 \includegraphics[width=12cm,clip]{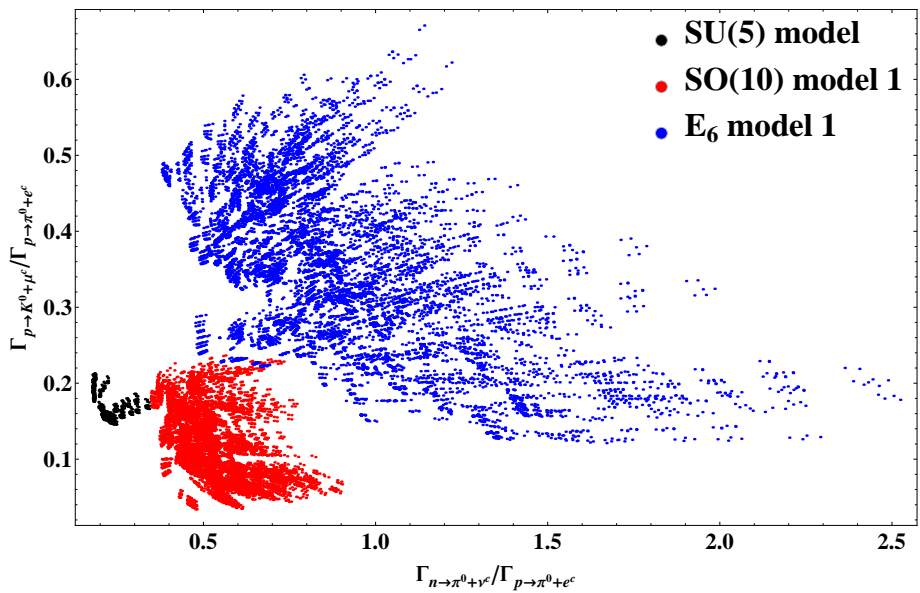}
 \caption{In $SU(5)$ we have $10^3$ model points, because we have three independent 
 diagonalizing matrices and we generate 10 unitary matrices for each independent matrix.
 In $SO(10)$ and $E_6$ we have $10^4$ model points, because we have four independent 
 diagonalizing matrices.
 VEVs are taken as $x=1\times 10^{16}$ GeV, $v_c=5\times 10^{14}$ GeV,
 and $v_{\phi}=5\times 10^{15}$ GeV.}
 \label{Identification1}
\end{figure}

In order to distinguish the $SO(10)$ and $E_6$ models, we propose an additional
ratio of partial decay widths, 
$R_2\equiv\frac{\Gamma_{p \rightarrow K^0 + \mu^c}}{\Gamma_{p \rightarrow \pi^0 + e^c}}$ . One important fact is that the $SO(10)$ superheavy
gauge boson $X'$ cannot induce the effective interactions which include the second
generation fields which come from $\bf 10$ of $SO(10)$. On the other hand, the 
$E_6$ superheavy gauge boson $X''$ induces only the effective interactions which
include the second generation fields from $\bf 10$ of $SO(10)$. Therefore, the ratio
$\frac{\Gamma_{p \rightarrow K^0 + \mu^c}}{\Gamma_{p \rightarrow \pi^0 + e^c}}$ 
can play an important role in identifying the grand unification group.
See Figure \ref{Identification1}. We plot 
$R_1$ 
on the horizontal axis and 
$R_2$ 
on the vertical axis.
The figure shows that various model points can be classified into three regions
corresponding to the three grand unification groups, $SU(5)$, $SO(10)$, and $E_6$. 
These three GUT classes can be distinguished by these observations.

Of course, these results are strongly dependent on the explicit models and their 
parameters, especially
the VEVs, which we have taken as $x=1\times 10^{16}$ GeV, $v_c=5\times 10^{14}$ GeV,
 and $v_\phi=5\times 10^{15}$ GeV. However, we should note that the effect of
 $SO(10)$ superheavy gauge boson $X'$ is almost maximal in these VEVs because
 $v_c<<x$. On the other hand, the contribution from the $E_6$ superheavy
 gauge boson $X''$ can be larger because the contributions to the $X''$ mass 
 from the VEV $v_\phi$ and from the VEV $x$ are comparable in these parameters. 
 Therefore, if the ratio $R_1$ is observed to be much larger than one, the
observation suggests $E_6$ gauge group strongly.

\begin{figure}[htb]
 \centering
 \includegraphics[width=12cm,clip]{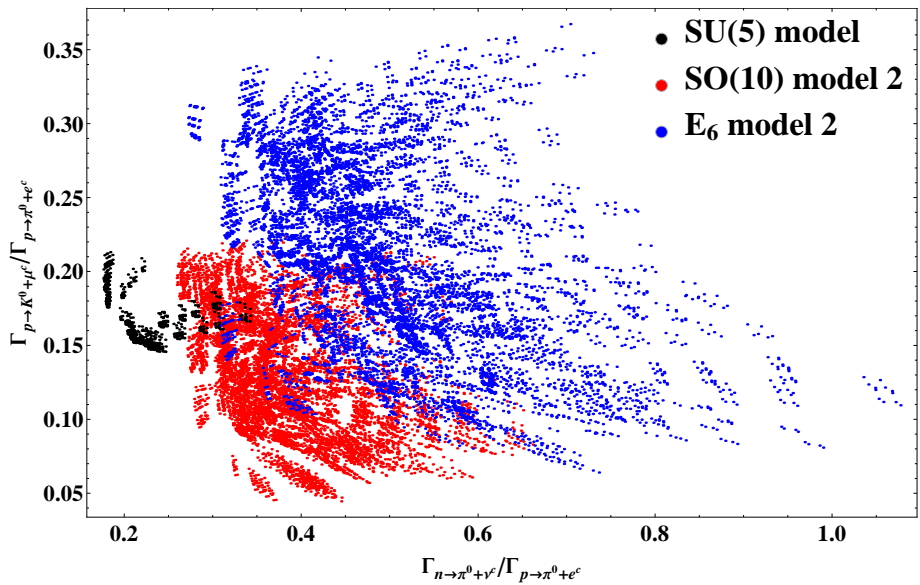}
 \caption{In $SU(5)$ we have $10^3$ model points, in $SO(10)$ and $E_6$ we have $10^4$ model 
 points, as noted in figure caption of Figure \ref{Identification1}.
 VEVs are taken as $x=v_c=v_{\phi}$.}
 \label{Identification2}
\end{figure}

If anomalous $U(1)_A$ symmetry is not adopted, usually the VEV relations 
$v_c, v_\phi\geq x$ are required in order to explain the gauge coupling unification.
Of course, if $v_c, v_\phi>>x$, then the predictions of $SO(10)$ models
and $E_6$ models become the same as those of $SU(5)$ models.
Here, we show another plot by taking $x=v_c=v_{\phi}$, which makes the 
$X'$ and $X''$ contribution maximal in these models without anomalous $U(1)_A$ symmetry, 
keeping the success of the gauge coupling unification.
The results are shown in Figure \ref{Identification2}. It is understood that
the $SO(10)$ model points come closer to the $SU(5)$ model points and the
$E_6$ model points come closer to the $SO(10)$ model points.

In the last of this subsection, we will explain why we adopt
$n\rightarrow \pi^0 + \nu^c$ mode instead of the $p\rightarrow \pi^+ + \nu^c$ mode.
We have two reasons.  
First, the former mode is easier to be detected experimentally.
Since the decay of $\pi^+$ includes an invisible neutrino, the latter decay mode
is more difficult to be observed.
The other reason is that the hadron matrix element of the 
former mode is the same as that of $p \rightarrow \pi^0 + e^c$ mode, and therefore 
in the ratio $R_1$ these hadron matrix elements are cancelled.

\section{Discussion and Summary}
We have calculated the lifetime of the nucleon for various decay modes via dim.
6 operators in the anomalous $U(1)_A$ GUT models. 
Since the anomalous $U(1)_A$ GUT
models predict lower unification scale in general, it is important to predict the 
nucleon lifetime via dim. 6 operators. The lifetime 
$\tau(p\rightarrow \pi^0 + e^c)$ has been calculated as $O(10^{34})$ years for 
the unification scale $\Lambda_u=1\times 10^{16}$ GeV, which is a typical value
for the unification scale in anomalous $U(1)_A$ GUT scenario with the $U(1)_A$ 
charge of the adjoint Higgs $a=-1/2$. Although we have several ambiguities in the
calculation from $O(1)$ coefficients or the hadron matrix elements, the discovery 
of the nucleon decay in next experiments\cite{Hyper-Kamiokande} can be expected because the present experimental 
lower limit is $1.29\times 10^{34}$ years. 
The predicted value can become $O(10^{33})$ years for the anomalous $U(1)_A$ GUT 
models with $a=-1$. In the calculation, we have taken into account the ambiguities from the quark and
 lepton mixings by generating the various diagonalizing unitary matrices randomly. 
One of the largest ambiguities for the predictions comes from the $O(1)$ coefficient
of the unification scale. 
Since the lifetime is proportional to $\Lambda_u^4$, the 
factor 2 in the unification scale can make the prediction of 
the lifetime 16 times larger.
 Moreover,
the ambiguities from the hadron matrix elements can easily change the prediction 
by factor 2. Therefore, we cannot reject the anomalous $U(1)_A$ GUT with $a=-1$
by these predictions. 
We can expect the observation of the nucleon decay in near future
experiments.

We have proposed that the two ratios, 
$R_1\equiv\frac{\Gamma_{n \rightarrow \pi^0 + \nu^c}}{\Gamma_{p \rightarrow \pi^0 + e^c}}$ and
$R_2\equiv\frac{\Gamma_{p \rightarrow K^0 + \mu^c}}{\Gamma_{p \rightarrow \pi^0 + e^c}}$,
are important to identify the anomalous $U(1)_A$ GUT models. 
The ratio $R_1$ becomes larger for the larger rank of the grand unification group if
the masses of the $SO(10)$ and $E_6$ superheavy gauge bosons $X'$ and $X''$
are comparable or even smaller than the $SU(5)$ superheavy gauge boson mass.
This is because the superheavy gauge bosons $X'$ and $X''$ induce only the effective
 interactions which include the doublet lepton $l$,
while the $SU(5)$ superheavy gauge boson $X$ induces both the effective interactions 
with $l$ and the effective interactions with $e_R^c$.  
What is important is that in the anomalous $U(1)_A$ GUT models, the $X'$ mass is always 
comparable with the $X$ mass. The $X''$ mass can be smaller than the $X$ mass, that
is dependent on the explicit models. Therefore, at least in the anomalous $U(1)_A$
GUT scenario, measuring this ratio is critical in distinguishing the $SU(5)$
models from the other models.
The ratio $R_2$ is important to distinguish $E_6$ models from $SO(10)$ models. 
In most of the anomalous $U(1)_A$ GUT models with $SO(10)$ and $E_6$ unification group,
the $\bf\bar 5'$ field from $\bf 10$ of $SO(10)$ becomes the main component of the
second generation $\bf\bar 5$ field to obtain large neutrino mixings.
What is important here is that the $X'$ boson does not induce the effective 
interactions which include $\bf\bar 5'$ fields, while the $X''$ boson induces
only the effective interactions which include $\bf\bar 5'$. Therefore, in $E_6$
unification, the nucleon decay widths for the second generation quark and lepton
must be larger than in $SO(10)$ unification.
We have plotted various model points in several figures in which the horizontal axis
is $R_1$ and the vertical axis is $R_2$. And we have concluded that
we can identify the grand unification group by measuring these ratios if 
$x=1\times 10^{16}$ GeV, $v_c=5\times 10^{14}$ GeV, and $v_\phi=5\times 10^{15}$ 
GeV, which are typical values in the models with $a=-1/2$, $c+\bar c=-5$, and
$\phi+\bar\phi=-2$. 
 Of course, this conclusion is dependent on the parameters. 
For example, 
when $v_\phi>>x$, it becomes difficult to distinguish the $E_6$ models from
the $SO(10)$ models because the mass of $X''$ becomes much larger than the other
superheavy gauge bosons.
However, since it is difficult to realize $R_1>0.4$ in $SU(5)$ unification, 
if $R_1$ is observed to be larger than 0.4, then the grand unification group is not
$SU(5)$. Moreover, if $R_2$ is larger than 0.3, $E_6$ unification is 
implied. An important point is that $\Gamma(n\rightarrow \pi^0+\nu^c)$ and 
$\Gamma(p\rightarrow K^0+\mu^c)$ can
be comparable with $\Gamma(p\rightarrow \pi^0+e^c)$ in $E_6$
unification.

Note that our calculations can apply to the usual SUSY GUT models in which the 
unification scale is around $\Lambda_G=2\times 10^{16}$ GeV, although the predicted 
lifetime becomes much longer. And taking account of the gauge coupling unification,
the VEVs $v_c$ and $v_\phi$ must be larger than $\Lambda_G$ usually. Therefore,
the effects of superheavy gauge bosons $X'$ and $X''$ are not so large.
However, the ratios $R_1$ and $R_2$ must be important in identifying GUT models
even without anomalous $U(1)_A$ gauge symmetry.

\section{Acknowledgement}
We thank J. Hisano and Y. Aoki to tell us the present status on the lattice calculation
of hadron matrix elements. 
N.M. is supported in part by Grants-in-Aid for Scientific Research from MEXT of 
Japan. This work is partially supported by the Grand-in-Aid for Nagoya University
Leadership Development Program for Space Exploration and Research Program from the MEXT 
of Japan.


\begin{thebibliography}{99}
\bibitem{GUT}
  H.~Georgi and S.~L.~Glashow,
  Phys.\ Rev.\ Lett.\  {\bf 32}, 438 (1974).

\bibitem{anarchy}
  L.~J.~Hall, H.~Murayama and N.~Weiner,
  Phys.\ Rev.\ Lett.\  {\bf 84}, 2572 (2000)
  [hep-ph/9911341].

  J.~Hisano, K.~Kurosawa and Y.~Nomura,
  Nucl.\ Phys.\ B {\bf 584}, 3 (2000)
  [hep-ph/0002286].

\bibitem{nucleon decay dim6}
  H.~Georgi, H.~R.~Quinn and S.~Weinberg,
  Phys.\ Rev.\ Lett.\  {\bf 33}, 451 (1974).

\bibitem{nucleondecayoperator}
   S.~Weinberg,
  Phys.\ Rev.\ Lett.\  {\bf 43}, 1566 (1979).
  
    L.~F.~Abbott and M.~B.~Wise,
  Phys.\ Rev.\ D {\bf 22}, 2208 (1980).

\bibitem{nucleon decay dim5}
  N.~Sakai and T.~Yanagida,
  Nucl.\ Phys.\ B {\bf 197}, 533 (1982).

\bibitem{SuperK}
  H.~Nishino {\it et al.}  [Super-Kamiokande Collaboration],
  Phys.\ Rev.\ D {\bf 85}, 112001 (2012)
  [arXiv:1203.4030 [hep-ex]].

\bibitem{dim5bound}
  T.~Goto and T.~Nihei,
  Phys.\ Rev.\ D {\bf 59}, 115009 (1999)
  [hep-ph/9808255].

\bibitem{dim5bound'}
  J.~Hisano, H.~Murayama and T.~Yanagida,
  Nucl.\ Phys.\ B {\bf 402}, 46 (1993)
  [hep-ph/9207279].

\bibitem{DTsplitting review}
For the review,
  L.~Randall and C.~Csaki,
  In *Palaiseau 1995, SUSY 95* 99-109
  [hep-ph/9508208].

\bibitem{DTsplitting}
  N.~Maekawa,
  Prog.\ Theor.\ Phys.\  {\bf 106}, 401 (2001)
  [hep-ph/0104200].

\bibitem{GCUA}
  N.~Maekawa,
  Prog.\ Theor.\ Phys.\  {\bf 107}, 597 (2002)
  [hep-ph/0111205].

\bibitem{GCUA'}
  N.~Maekawa and T.~Yamashita,
  Phys.\ Rev.\ Lett.\  {\bf 90}, 121801 (2003)
  [hep-ph/0209217].
  
\bibitem{E6LLM}
  M.~Bando and N.~Maekawa,
  Prog.\ Theor.\ Phys.\  {\bf 106}, 1255 (2001)
  [hep-ph/0109018].

\bibitem{E6Higgs}
  N.~Maekawa and T.~Yamashita,
  Prog.\ Theor.\ Phys.\  {\bf 107}, 1201 (2002)
  [hep-ph/0202050].

\bibitem{missing partner}
  H.~Georgi,
  Phys.\ Lett.\ B {\bf 108}, 283 (1982).

  A.~Masiero, D.~V.~Nanopoulos, K.~Tamvakis and T.~Yanagida,
  Phys.\ Lett.\ B {\bf 115}, 380 (1982).

  B.~Grinstein,
  Nucl.\ Phys.\ B {\bf 206}, 387 (1982).

\bibitem{formfactor}
  Y.~Aoki, C.~Dawson, J.~Noaki and A.~Soni,
  Phys.\ Rev.\  D {\bf 75}, 014507 (2007)
  [arXiv:hep-lat/0607002].

  Y.~Aoki, E.~Shintani and A.~Soni,
  arXiv:1304.7424 [hep-lat].

\bibitem{E6}
  F.~Gursey, P.~Ramond and P.~Sikivie,
  Phys.\ Lett.\ B {\bf 60}, 177 (1976).

  Y.~Achiman and B.~Stech,
  Phys.\ Lett.\ B {\bf 77}, 389 (1978).

  R.~Barbieri and D.~V.~Nanopoulos,
  Phys.\ Lett.\ B {\bf 91}, 369 (1980).

\bibitem{SU(3)*3'}
  M.~Bando and T.~Kugo,
  Prog.\ Theor.\ Phys.\  {\bf 101}, 1313 (1999)
  [arXiv:hep-ph/9902204].
  
  M.~Bando, T.~Kugo and K.~Yoshioka,
  Prog.\ Theor.\ Phys.\  {\bf 104}, 211 (2000)
  [hep-ph/0003220].

\bibitem{SU(3)*3}
  T.~W.~Kephart and M.~T.~Vaughn,
  Annals Phys.\  {\bf 145}, 162 (1983).

\bibitem{DW form}
  S.~Dimopoulos and F.~Wilczek,NSF-ITP-82-07

  M.~Srednicki,
  Nucl.\ Phys.\  B {\bf 202}, 327 (1982).

\bibitem{FNLLM}
  C.~D.~Froggatt and H.~B.~Nielsen,
  Nucl.\ Phys.\ B {\bf 147}, 277 (1979).

  L.~E.~Ibanez and G.~G.~Ross,
  Phys.\ Lett.\ B {\bf 332}, 100 (1994)
  [hep-ph/9403338].

\bibitem{CKM}
  M.~Kobayashi and T.~Maskawa,
  Prog.\ Theor.\ Phys.\  {\bf 49}, 652 (1973).

\bibitem{MNS}
  Z.~Maki, M.~Nakagawa and S.~Sakata,
  Prog.\ Theor.\ Phys.\  {\bf 28}, 870 (1962).

\bibitem{RFL}
  A.~J.~Buras, J.~R.~Ellis, M.~K.~Gaillard and D.~V.~Nanopoulos,
  \\
  Nucl.\ Phys.\  B {\bf 135}, 66 (1978).

\bibitem{RFS1}
  L.~E.~Ibanez and C.~Munoz,
  Nucl.\ Phys.\  B {\bf 245}, 425 (1984).

\bibitem{RFS2}
  C.~Munoz,
  Phys.\ Lett.\  B {\bf 177}, 55 (1986).

\bibitem{PDG}
  J.~Beringer {\it et al.}  [Particle Data Group Collaboration],
  Phys.\ Rev.\ D {\bf 86}, 010001 (2012).

\bibitem{PDGc}
pdgLive http://pdg8.lbl.gov/rpp2013v2/pdgLive/Viewer.action

\bibitem{theta13}
  F.~P.~An {\it et al.}  [DAYA-BAY Collaboration],
  Phys.\ Rev.\ Lett.\  {\bf 108}, 171803 (2012)
  [arXiv:1203.1669 [hep-ex]].

  Y.~Abe {\it et al.}  [DOUBLE-CHOOZ Collaboration],
  Phys.\ Rev.\ Lett.\  {\bf 108}, 131801 (2012)
  [arXiv:1112.6353 [hep-ex]].

  J.~K.~Ahn {\it et al.}  [RENO Collaboration],
  Phys.\ Rev.\ Lett.\  {\bf 108}, 191802 (2012)
  [arXiv:1204.0626 [hep-ex]].

\bibitem{neutrinoSUM}
  P.~Fileviez Perez,
  Phys.\ Lett.\ B {\bf 595}, 476 (2004)
  [hep-ph/0403286].

  I.~Dorsner and P.~Fileviez Perez,
  Phys.\ Lett.\ B {\bf 605}, 391 (2005)
  [hep-ph/0409095].

  I.~Dorsner, S.~Fajfer and N.~Kosnik,
  Phys.\ Rev.\ D {\bf 86}, 015013 (2012)
  [arXiv:1204.0674 [hep-ph]].
  
\bibitem{SU5 VS SO10}
  F.~Wilczek and A.~Zee,
  Phys.\ Rev.\ Lett.\  {\bf 43}, 1571 (1979).
  
  P.~Langacker,
  Phys.\ Rept.\  {\bf 72}, 185 (1981).

\bibitem{Hyper-Kamiokande}
  K.~Abe, T.~Abe, H.~Aihara, Y.~Fukuda, Y.~Hayato, K.~Huang, A.~K.~Ichikawa and M.~Ikeda {\it et al.},
  arXiv:1109.3262 [hep-ex].
  
\end{thebibliography}
\end{document}